\documentclass[structabstract,letter]{aa} 
\usepackage{graphicx}
\usepackage{txfonts}
\begin{document}

    \title{The diffuse supernova neutrino background: 
    Expectations and uncertainties derived from SN1987A}

    \author{F. Vissani \and G. Pagliaroli}

    \institute{INFN, Laboratori Nazionali del Gran Sasso, Assergi (AQ) Italy\\
              }
\date{}

\abstract
{The detection of the diffuse supernova neutrino background may be imminent,
but theoretical predictions are affected by substantial
uncertainties.}
{We calculate the signal and its uncertainty with the present configuration
of Super-Kamiokande and consider the possibility of lowering the threshold 
by means of gadolinium loading.}
{We model neutrino emission
following the analysis of SN1987A by Pagliaroli and collaborators 2009  
and use the number of expected
events in the neutrino detector as a free parameter of the fit. 
The best-fit value of this parameter and 
its error are evaluated by means of standard maximum likelihood procedures, 
taking into account properly the correlations.}
{The uncertainties in the astrophysics of the emission dominates the total
uncertainty in the expected signal rate, which conservatively ranges 
from $0.3$ to $0.9$ events per year and from $1.1$ to $2.9$ with gadolinium.} 
{}

\titlerunning{DSNB expectations and uncertainties from SN1987A}
\maketitle


\section{Introduction}

Massive stars end their life exploding as core 
collapse supernovae (SN) and leaving compact remnants, such as neutron
stars or black holes. Most of the binding energy 
is released by means of neutrino emission, though the details of the
emission are not yet completely understood.
The SN1987A event provided evidence that the neutrino emission exists and is detectable
(Hirata \emph{et al.} \ \cite{Hirata}; Bionta \emph{et al.}\
\cite{IMB}; Alekseev \emph{et al.}\ \cite{baksan}). The neutrino signal offers us the most
promising opportunity to date to probe this unique astrophysical system.

Neutrino emission from core-collapse supernovae that have exploded in the Universe 
cumulate in the diffuse supernova neutrino background
(DSNB). The prospects of detection seem promising; for 
reviews and references to the original papers, we refer  Ando \& Sato
(\cite{Ando}) and Beacom (\cite{Beacom}). 
To predict the DSNB signal, 
two different quantities are needed: the explosion
rate of core collapse SNe as a function of the redshift and the average
neutrino emission of the individual supernovae. Each of these
quantities is known up to an error which implies uncertainty in the
prediction. Until both uncertainties are quantified, it is impossible to 
evaluate the reliability of the expectations.

The rate of core-collapse supernovae can be identified with the
rate of massive-star formation as a function of redshift.
This quantity, in turn, can be obtained from knowledge of the cosmic history of
star formation and the initial distribution of mass. 
Horiuchi \emph{et al.} (\cite{Horiuchi:2008jz})
used a comprehensive compilation of data to evaluate the 
uncertainty in the rate.

We attempt is to complement the study of the uncertainties by evaluating the impact of 
the neutrino emission for an individual supernova. Following Fukugita \& Kawasaki
(\cite{Fukugita}), we base our inferences on SN1987A observations using
for the analysis the neutrino emission model of Pagliaroli~\emph{et al.} (2009).
We evaluate the expected signal and its uncertainty  for the Super-Kamiokande detector. We consider its present configuration, 22.5 kton of fiducial volume with 
a threshold of $E_{th}=19.3$ MeV, Malek et al.\ (\cite{Malek})  
and consider the possibility of lowering the threshold to 11.3 MeV by employing gadolinium loading as advocated by Beacom \& Vagins (\cite{bv}).

\section{The emission from an individual supernova}
\label{sec1}

We determine the $\bar\nu_e$ spectrum from SN1987A data,
which we then use for the analysis of DSNB. This is done by
parameterizing the time and energy distribution,  then by fitting the data from Kamiokande-II, IMB, and Baksan
and finally by integrating over the time distribution.
The main motivation behind this procedure is that we still do not have
a definitive theory of supernova explosion; 
thus, despite their paucity, the data from SN1987A play
a very important role in guiding our understanding of supernova neutrino manifestations. 

\paragraph{The model for supernova emission}

Motivated by the prospect of exploring, through the neutrino signal,  
the physics and astrophysics of the gravitational collapse, we proposed in 
Pagliaroli \emph{et al.} (\cite{AP}) a parameterization of
$\bar{\nu}_e$ emission, based on our present understanding of emission
processes improving the model by Loredo \& Lamb (\cite{ll}). 
Our model has two emission phases and the $\bar\nu_e$ flux is
\begin{equation}
\Phi_{\bar\nu_e}(t,E') = 
\Phi_{a}(t,E') + [1-j_k(t)]\ \Phi_{c}
(t-\tau_a,E'), \label{flux}
\end{equation}
where $t$ is the emission time and $E'$ is the emitted neutrino energy. 
The first term $\Phi_{a}$ is the flux
generated during the phase of accretion and 
above the shock  by the interactions between the positrons and the target 
neutrons. It describes volume emission. It uses three free parameters: the
initial accreting mass $M_a$, the time scale $\tau_a$, and the initial
temperature of the positrons $T_a$. The second term $\Phi_{c}$ is the flux coming from the
thermal emission of the new-born proto-neutron star (i.e., the cooling
phase). It describes radiation from a surface, whose area is
proportional to the radius of the neutrino sphere $R_c$, 
with a time scale $\tau_c$ and the initial temperature of the emitted
antineutrinos $T_c$. Finally, the function $j_k(t)$ links smoothly the two
emission phases, delaying the cooling emission by $\tau_a$.
Analytic expressions of these three functions,
$\Phi_{a}$, $ \Phi_{c}$, and $j_k(t)$ are given in Eqs.~10, ~13
and ~18 of Pagliaroli \emph{et al.} (\cite{AP}), respectively.
We include neutrino oscillations with normal hierarchy as discussed in
Sec. C of Pagliaroli \emph{et al.} (\cite{AP}).

\paragraph{SN1987A data analysis}

In Pagliaroli \emph{et al.} (\cite{AP}), we tested the parameterized 
model using the small set of events collected in 1987 by Kamiokande-II, IMB, and Baksan,
leaving aside an interpretation of the events recorded by LSD
(Aglietta \emph{et al.} \cite{LSD}). 
When we compared the fit based only on the cooling phase, the one  
adopted in usual SN1987A data analyses, with our fit, we  found that 
the two-phase emission model is 50 times more probable (i.e., 
we got a $2.5 \sigma$ indication in its favor).  Moreover, the best-fit values
of the astrophysical parameters, namely 
$R_c=16 \mbox{ km}$, $M_a=0.22 M_{\odot}$, $T_c=4.6 \mbox{ MeV}$, $T_a=2.4 \mbox{
MeV}$, $\tau_c=4.7 \mbox{ s}$, and $\tau_a=0.55 \mbox{ s}$,  agree well
with the general expectations. For instance, the duration
of the accretion phase is shorter than one second, 
the radius of the neutrino sphere is similar in size to the neutron
star, and the total radiated energy 
$2.2\times 10^{53}$ erg is similar to the binding energy.
Finally, the luminosity curve and the mean energy as 
functions of time both resemble the results of numerical
simulations. 
The errors in the parameters and their correlations 
are large, as expected from the limited number of events (see again
Pagliaroli \emph{et al.} \cite{AP}).

The previous analyses of SN1987A data of Lunardini
(\cite{Lunardini}) and Yuksel and Beacom (\cite{Yuksel}) designed 
to predict DSNB focused only on the energy
spectrum, motivated by the opinion that the time distribution of the events is irrelevant.
However, we show that our detailed theoretical description of neutrino
emission leads to a peculiar integrated spectrum that retains an
imprint of the two different emission phases.

\paragraph{The emission spectrum}

\begin{figure}
\includegraphics[width=7.0cm,angle=0]{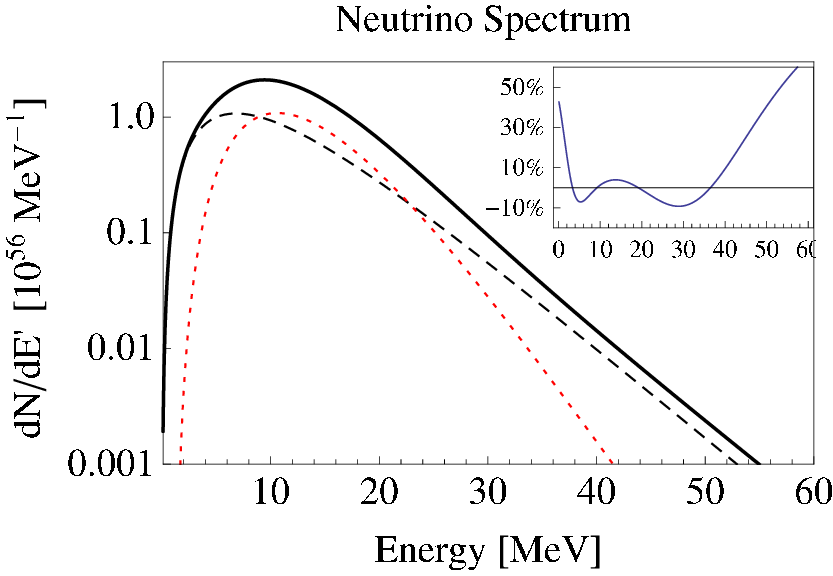}
\caption {The SN1987A best-fit spectrum (continuous line) and the two
 contributions from accretion phase (dotted line) and cooling phase
 (dashed line). In the inset, we plot the percentage difference with
 the Fermi-Dirac approximation of the spectrum, given in the text.} \label{spettro}
\end{figure}

By substituting in the $\bar{\nu}_e$ flux of
Eq.(\ref{flux}) the best-fit values reported in the previous paragraph
and integrating both in a $30$ s time window and on the
emitting surface, we obtain the reference spectrum $dN/dE'$ for a
single supernova event from SN1987A shown in Fig.\ \ref{spettro}.
We compare this spectrum with a Fermi-Dirac distribution, i.e., 
$\frac{k}{T^3}\frac{E'^2}{1+\exp(E'/T-\eta)}$. To obtain the same
total integral and the same first two momenta between 
our spectrum and the Fermi-Dirac distribution, we need $k =2.97\times 10^{57}$,
a temperature $T = 3.76 \mbox{MeV}$, and a pinching factor $\eta =
0.679$. However, the percentage difference between the Fermi-Dirac spectrum and our
spectrum is not negligible as shown in the inset of Fig.\
\ref{spettro}. 
The agreement worsens using the parametrization
of Keil {\em et al.}\ (\cite{keil}).
The emitted electron antineutrinos of our best-fit spectrum are of relatively low
energy; half of them are emitted 
at energies below 11.2 MeV. In particular, 
when we calculate the number of events $N_{ev}$ expected from a galactic
supernova assuming an energy threshold of $E_{min}>6.5$ MeV, 
$N_{ev}\propto \int dE' \sigma(E) \Phi_{\bar\nu_e}$ as a function of the 
upper extreme of integration $E_{max}$,
we find that there are the 25\%, 50\%, and 75\% of the events below 
$E_{max}$=14, 18, and 24 MeV, respectively.\footnote{Of course, 
$\sigma$ denotes the cross-section for the IBD process $\bar\nu_e p\to
e^+ n$. We use Eq.~25 of Strumia \& Vissani (\cite{Strumia:2003zx}) 
and throughout the paper, we adopt the approximation $E_{\bar\nu_e}=E_{e^+}+1.3$ MeV, 
which is adequate for our needs. Unless specified otherwise, we always refer to
$\bar\nu_e$ energy.}

\section{Expectations for the diffuse neutrino flux}
We estimate the diffuse neutrino flux  accumulated by all the past supernovae
exploded in the Universe, assuming that the antineutrino flux discussed 
in the previous Section represents the typical emission 
of a core-collapse event. 

In a Friedmann-Robertson-Walker flat Universe, 
the expected DSNB flux is
\begin{equation}
\frac{d\phi(E)}{dE}=\frac{c}{H_0}\int_{0}
dz
\frac{R_{CCSN}(z)}{\sqrt{\Omega_\Lambda+\Omega_m(1+z)^3}}\frac{dN(E')}{dE'}.
\label{main}
\end{equation}
The last term of the integrand, $dN/dE'$, is the spectrum of a single
SN emission discussed above and calculated for the red-shifted
energy $E' =(1+z)E$. The Hubble constant is $H_0=71 \mbox{ km }
\mbox{s}^{-1} \mbox{Mpc}^{-1}$, $c$ is the light speed, and the
values of the cosmological constants are $\Omega_m =0.27$,
$\Omega_\Lambda= 0.73$ as measured from the WMAP
experiment in Jarosik \emph{et al.} (\cite{Jarosik:2010iu}).  These assumptions do not introduce 
significant errors into the predictions.

\begin{figure}
\includegraphics[width=8.5cm,angle=0]{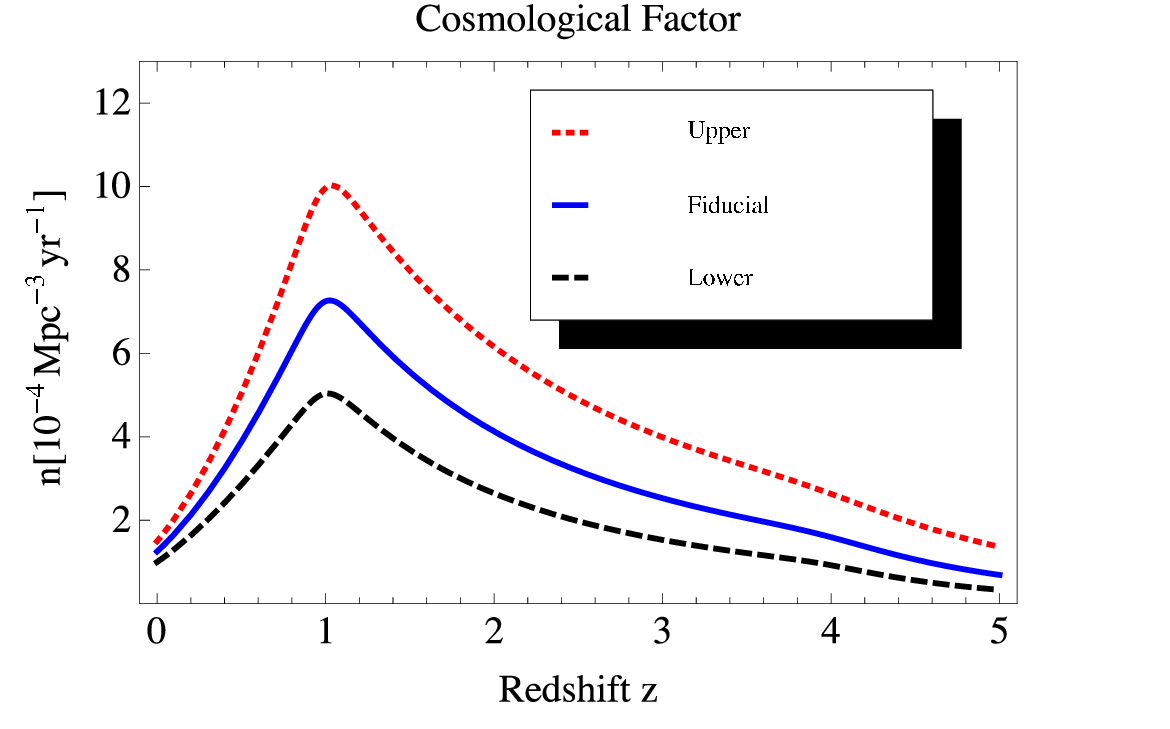}
\caption {The cosmological factor $n$ of Eq.~(\ref{nterm})
for various fits of the core-collapse rate. The continuous line is
for the fiducial rate, the dotted line is for the upper one, and
the dashed line is for the lower rate.} \label{cosmology}
\end{figure}

The key term we have to discuss is $R_{CCSN}(z)$, namely the rate of core-collapse SN for comoving
volume as a function of the redshift $z$. This function can be obtained from 
the number of stars formed in a comoving volume, i.e., from the
star formation rate as a function of the redshift, and from the
fraction of stars in the appropriate mass range for core collapse SNe.
Following Horiuchi \emph{et al.} (\cite{Horiuchi:2008jz}), we adopt the 
initial mass function of Salpeter (\cite{Salpeter})
and three different analytic fits for the rate of core-collapse to
take into account the astrophysical uncertainty.  
In Fig.(\ref{cosmology}), we show the quantity 
\begin{equation}
n(z)=
\frac{R_{CCSN}(z)}{\sqrt{\Omega_\Lambda+\Omega_m(1+z)^3}},
\label{nterm}
\end{equation} 
which we call the ``cosmological
factor'', namely the first integrand term in Eq.~(\ref{main}). 
This plot summarizes all the relevant 
assumptions about the cosmology and the distribution of the sources.
Horiuchi \emph{et al.} (\cite{Horiuchi:2008jz})
demonstrated that the use of either the Kroupa (\cite{kroupa})
or Baldry \& Glazebrook (\cite{BG}) initial mass functions,
or the uncertainties in the lowest mass that forms a supernova, 
introduces only a negligible error in the estimates.

\begin{figure}
\includegraphics[width=7.5cm,angle=0]{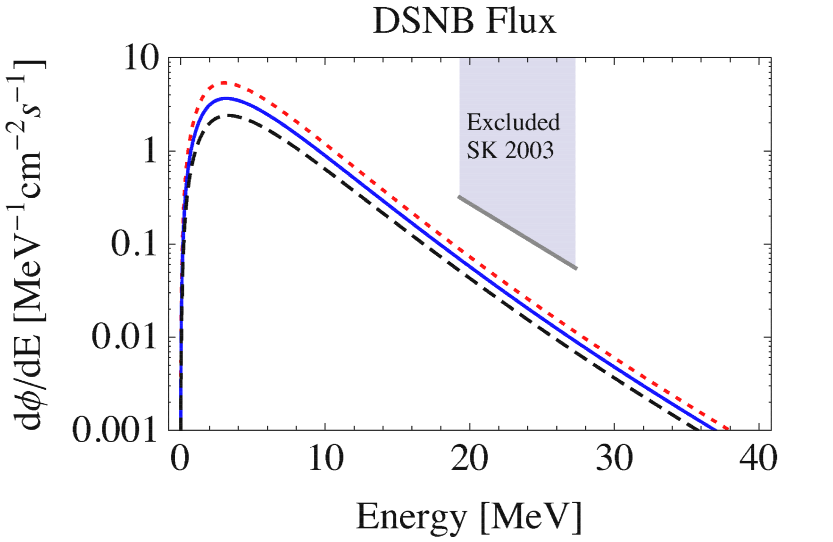}
\caption {Expected DSNB flux in a logarithmic scale for the three different fits of
core-collapse rate. Also is shown the upper limit of this quantity given by Super-Kamiokande in
2003 in the energy range between $19.3$ and $27.3$ MeV.
The meaning of continuous, dotted and dashed lines as in Fig.~(\ref{cosmology}).}
\label{DSNBflux}
\end{figure}

We now use Eq.~(\ref{main}) to calculate 
the flux of the diffuse SN neutrino background
$d\phi/dE$
as a function of the neutrino energy at the
Earth.
We plot it in Fig.~(\ref{DSNBflux}) along with the upper bound in the
energy region $E>E_{th}=19.3$ MeV of $1.2\ \bar{\nu}_e$ cm$^{-2}$ s$^{-1}$
obtained by the Super-Kamiokande collaboration (Malek \cite{Malek}).
The total flux of DSNB with the SN1987A best-fit model is $\phi= 27.2
\mbox{ cm}^{-2}\mbox{ s}^{-1}$ for 
the fiducial rate. 
Only $\sim1\%$ of this flux can be actually
observed with a neutrino energy threshold of
$19.3$ MeV; with the use of gadolinium, 
the accessible spectrum becomes $\sim 8\%$, showing that most of the flux 
falls in the very low energy region.

\paragraph{Event rate in Super-Kamiokande\label{ss3}}

%

We consider the events expected in a detector
such as Super-Kamiokande, with a fiducial mass of $M_d=22.5$
kton of water and a detection efficiency $\epsilon$ set 
equal to the $98\%$ above a neutrino energy threshold 
($E_{th}=19.3$ MeV in the present configuration or $E_{th}=11.3$ MeV by 
loading the detector with gadolium).
However, we note that a scintillator-based detector with
average chemical formula C$_9$H$_{21}$ (or  C$_6$H$_3$(CH$_3$)$_3$)
and with 16 (or 25) kton of mass has the same number of target protons $N_p=1.5\times 10^{33}$,
thus the same number of $\bar\nu_e p\to e^+ n$ interactions.
The event rate is calculated easily as  
\begin{equation}
\dot{N}_{ev}=N_p \int_{E_{th}} dE \sigma(E) \frac{d\phi}{dE} \epsilon(E) .
\label{trivia}
\end{equation}
For the best-fit model adopted, we expect to see in
Super-Kamiokande the number of events per year reported in
Table(\ref{Neventi}) for two different energy thresholds and three 
different fits of the rate. We expect $0.39-0.65$ events per year when we consider a threshold of $19.3$
MeV; this increases to $1.35-2.35$ events per year lowering the
energy threshold to $11.3$ MeV.
Our results agree well with those of Horiuchi \emph{et al.} (\cite{Horiuchi:2008jz}).

\begin{table}[b]
\caption{Rate of DSNB events expected in Super-Kamiokande detector [$\mbox{yr}^{-1}$]. 
Together with the best-fit values, we show the $1\sigma$ and
$2\sigma$ statistical errors obtained from marginalization
procedure of Sect.~\ref{secf}.} \label{Neventi}
\begin{tabular}{|c|c|c|}
  \hline
  CCSN Rate &  $E_{th}=19.3 \mbox{MeV} $  &  $E_{th}=11.3
  \mbox{MeV}$ \\ \hline \\[-2ex]
   Upper & $0.65^{+0.23}_{-0.20}(1\sigma)$ $^{+0.51}_{-0.34}(2\sigma)$ & $2.35^{+0.59}_{-0.51}(1\sigma)$
  $^{+1.24}_{-0.92}(2\sigma)$\\[1ex]
  Fiducial & $0.51^{+0.18}_{-0.16}(1\sigma)$ $^{+0.40}_{-0.26}(2\sigma)$ & $1.82^{+0.46}_{-0.39}(1\sigma)$ $^{+0.97}_{-0.71}(2\sigma)$ \\[1ex]
  Lower & $0.39^{+0.13}_{-0.12}(1\sigma)$ $^{+0.30}_{-0.20}(2\sigma)$ & $1.35^{+0.34}_{-0.29}(1\sigma)$ $^{+0.71}_{-0.52}(2\sigma)$ \\[1ex]
  \hline
\end{tabular}
\end{table}

\begin{figure}
\includegraphics[width=7.5cm,angle=0]{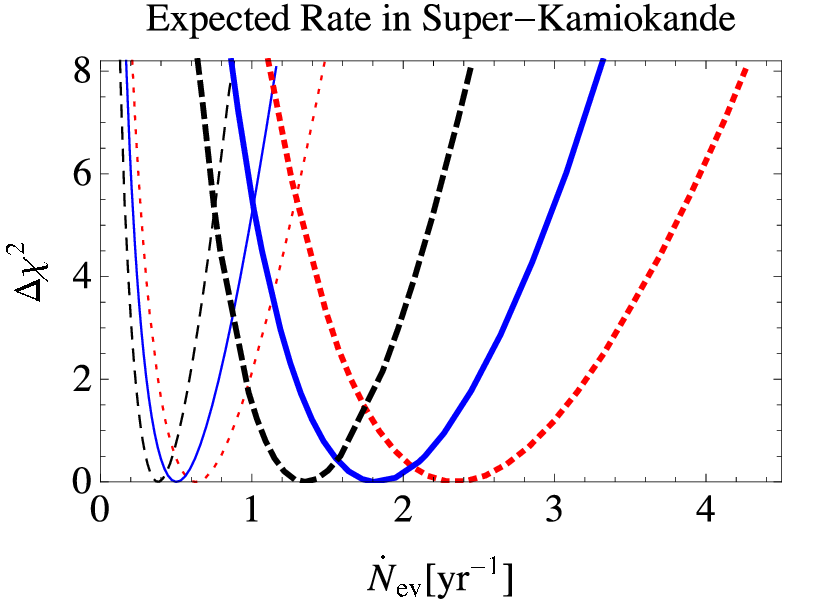}
\caption {$\Delta \chi^2$ curves as a function of event rate $\dot{N}_{ev}$ in
Super-Kamiokande, as obtained from SN1987A data analysis and for two
different energy thresholds. The set of thin lines correspond to the
present threshold of $19.3$ MeV whereas the set with thick lines
correspond to $E_{th}=11.3$ MeV. The meaning of the individual lines is as in the
previous figures.} \label{chi2}
\end{figure}

\section{Uncertainty due to the emission model\label{secf}}

In Sect.~\ref{ss3}, we discussed the expectations for the DSNB
flux based on the SN1987A best-fit model and we considered only
the uncertainty related to the core-collapse rate. 
However, the SN1987A data set is small and the procedure of 
using only the best-fit model is doubtful, until one estimates 
the associated theoretical error. The main aim of this work is, indeed, to quantify this 
uncertainty, which is an entirely new result.


We proceed as follows. The likelihood in the SN1987A data
analysis used by Pagliaroli \emph{et al.} (\cite{AP}) is a function of
six unknown astrophysical parameters. The same is true for the DSNB 
rate $\dot{N}_{ev}$. For a given energy threshold, this can be written explicitly as
$\dot{N}_{ev}=M_a\ f(T_a,\tau_a)+R_c^2\ g(T_c,\tau_c,\tau_a)$, 
where the two contributions 
correspond to the two terms of Eq.~(\ref{flux}) and 
$f$ and $g$ are known functions.
The idea is simply to rewrite the likelihood, 
substituting the neutrino-sphere radius $R_{c}$ 
with the quantity in which we 
are interested, $\dot{N}_{ev}$.
We can then apply  
a standard marginalization procedure: for each 
fixed value of $\dot{N}_{ev}$ and varying the other five parameters, 
we find the maximum value of 
the likelihood  $L$. Thus, 
we obtain $\Delta \chi^2=2 (\ln L_{max}-\ln L)$ a function 
of $\dot{N}_{ev}$ only.
 

%
\begin{figure}[t]
\includegraphics[width=7.5cm,angle=0]{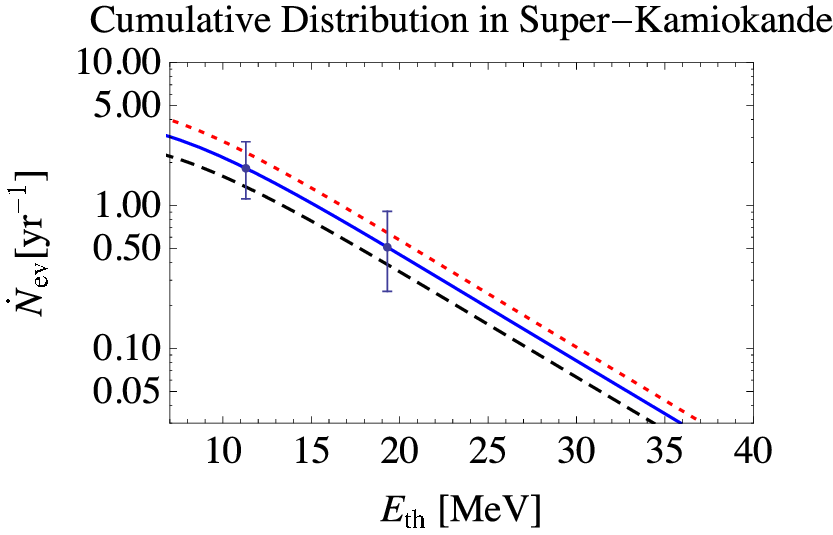}
\caption {Cumulative distribution of the events rate in
Super-Kamiokande, with the same
notation as Fig.~(\ref{cosmology}) for the lines. 
We give the $2\sigma$ error bars on that
prediction when $E_{th}=11.3$ MeV and $=19.3$ MeV.} \label{cum}
\end{figure}
%



The results of the numerical evaluations of
this $\Delta \chi^2$ are shown in Fig.(\ref{chi2}).
The thin lines represent the neutrino energy threshold 
used in Malek et al.\ (\cite{Malek}), $E_{th}=19.3 \mbox{MeV}$, 
whereas the thick ones correspond to the threshold of $E_{th}=11.3 \mbox{MeV}$ 
that can be achieved with gadolinium.
Three lines of each type are given, for the three 
different fits of the cosmic rate for core-collapse events. 
The $1\sigma$ and $2\sigma$ errors found from the $\Delta\chi^2$ 
curves are reported in Table (\ref{Neventi}).

We note that this procedure properly takes 
into account the correlations among the 
parameters, that are dictated by the likelihood function, without the 
need to explicitly calculate the covariance matrix.

\section{Summary and discussion}

We have evaluated the signal expected for diffuse supernova neutrino
background in Super-Kamiokande and its associated uncertainties, using
SN1987A data to constrain the model of neutrino emission. 
The results are summarized in Fig.(\ref{cum}), where we show the cumulative rate of DSNB
events in Super-Kamiokande as a function of the energy threshold
for the three different fits of the core-collapse rate. 

Following Horiuchi \emph{et al.} (\cite{Horiuchi:2008jz}), we used the
three different fits as a ``generous'' assessment of the cosmological uncertainty, 
although we are unable to attach to this range a precise statistical meaning.
The result is given in Table (\ref{Neventi}): the cosmological uncertainty is $\sim 25\%$ for the higher threshold 
and $\sim 27\%$ for the lower threshold, in reasonable agreement with
Horiuchi, despite the different emission model used.

On the other hand, selecting the fiducial cosmological model for the rate, we evaluated the
uncertainty on the DSNB due to the emission model. The $1\sigma$ (resp. $2\sigma$) percentage
error is $\sim 33\%$ (resp. $65\%$) for the higher threshold and
becomes $\sim 23\%$ (resp. $46\%$) for the lower threshold.
The prediction is more precise in the second case since the SN1987A data
provide a tighter constraint of the low energy region of the spectrum.
On the basis of the results of Pagliaroli \emph{et al.}
  (\cite{AP}), we conservatively estimate a 10\% systematic
uncertainty in our spectrum, which is intrinsic to the parameterization 
used in the fit.
Our $2\sigma$ ranges are curiously similar to those obtained by Ando \&
Sato (\cite{Ando}), who use three different emission models obtained from
numerical simulations. 

It is difficult to combine the two errors in a robust way, since only the one related to the emission model
has a precise statistical meaning. We thus construct a global range for the expected rate of DSNB
events by the following conservative procedure. The upper value of the range is obtained by
summing the best-fit value for the upper core collapse SN rate and its upward
$1\sigma$ (resp., $2\sigma$) statistical error; similarly, for the lower value of the range. 
In particular for $E_{th}=19.3$ MeV, the 
expected event rate in Super-Kamiokande ranges between $0.27$  
(resp., $0.19$) events per year and $0.88$ (resp., $1.16$) events per year,
giving a global uncertainty of $\sim 53\%$ (resp., $95\%$). For the lower
energy threshold $E_{th}=11.3$ MeV, the range becomes $1.06-2.94$
(resp., $0.83-3.59$) events per year with a percentage total error of
$47\%$ (resp., $76\%$). As quantified by the percentage errors, most of the global uncertainty
is due to the emission model uncertainty and to the low number of SN1987A data.

\appendix\section{An alternative expression for the rate}
We introduce an alternative expression for $\dot{N}_{ev}$ 
that emphasizes the role of the supernova flux at the source providing additional
insigh into the results.
{}From the emission spectrum 
$dN/dE'$, we
define an effective flux
\begin{equation}
\frac{d \Phi_*}{dE}(E')\equiv \frac{1}{4 \pi d^2 T}  \frac{dN}{dE'}
 \ \mbox{ with }\  \left\{
\begin{array}{l}
d\equiv 
\sqrt{ \frac{c}{4\pi n_* H_0 T}}=310\mbox{ kpc}  \\[1ex]
T\equiv 1\mbox{ yr}
\end{array}
\right.
\label{altr}
\end{equation}
where we have introduced the typical value for the
cosmic density of $n_*=2\times 10^{-4}$ SN/(Mpc$^3$ yr) and the observational
time $T$. 
By using as an integration variable $E'=E (1+z)$ (i.e., the neutrino energy at the emission),
we rewrite the signal rate in Eq.~(\ref{trivia})
mimicking closely the signal from a 
galactic supernova
\begin{equation}
\dot{N}_{ev}
=N_p \int_{E_{th}}^{\infty}\! dE' \sigma(E') \frac{d \Phi_*}{dE'}(E')
\epsilon_*(E').
\label{a2}
\end{equation}
The distribution of cosmic
supernovae is contained in the function 
\begin{equation}
\epsilon_*(E')\equiv \int^{\frac{E'}{E_{th}}-1}_0\!\!
\frac{dz}{\scriptstyle 1+z}\ 
\frac{n(z)\ \sigma({\scriptstyle E'/(1+z)}) }{n_*\ \sigma({\scriptstyle E'})}\  \epsilon({\scriptstyle E'/(1+z)}).
\end{equation}
This function leads to a severe cut in the rate of events given by Eq.(\ref{a2}). 
Above the energy threshold, this function can be approximated linearly as 
\begin{equation}
\epsilon_*(E')\propto  E'-E_{th},
\end{equation}
which corresponds to DSNB selecting the 
highest energy tail of the spectrum. 
This is important in connection with SN1987A.   
We have only limited information on the highest energy tail of the spectrum,
mostly thanks to IMB, as noted by 
Fukugita \emph{et al.} \cite{Fukugita}. We can constrain far more reliably 
the spectrum at lower energies, thanks to data from Kamiokande-II and Baksan. 
Thus, we expect, by 
lowering the threshold, not only 
the number of events due to DSNB to increase, 
but also the accuracy of the predictions 
based in SN1987A to improve, 
in agreement with Fig.~\ref{cum}.
We also note that the numerical value of $d$ in Eq.~(\ref{altr}) 
agrees with the DSNB signal being modest.

\end{document}